\documentclass[twocolumn,preprintnumbers,amsmath,amssymb]{revtex4}

\usepackage{mathrsfs}
\usepackage{amssymb}
\usepackage{graphicx}
\usepackage{dcolumn}
\usepackage{bm}
\usepackage{textcomp}
\usepackage{amsmath}
\usepackage[titletoc]{appendix}

\begin{document}
\title {Improved results for sending-or-not-sending twin-field quantun key distribution: breaking the absolute limit of repeaterless key rate }
\author{ Hai Xu$^{1}$, Zong-Wen Yu$ ^{1,3}$, Cong Jiang$ ^{1}$, Xiao-Long Hu$ ^{1}$
and Xiang-Bin Wang$ ^{1,2,4,5\footnote{Email Address: xbwang@mail.tsinghua.edu.cn}\footnote{Also at Center for Atomic and Molecular Nanosciences, Tsinghua University, Beijing 100084, China}}$}

\affiliation{ \centerline{$^{1}$State Key Laboratory of Low Dimensional Quantum Physics, Department of Physics,}
\centerline{Tsinghua University, Beijing 100084, China}
\centerline{$^{2}$ Synergetic Innovation Center of Quantum Information and Quantum Physics,}
\centerline{University of Science and Technology of China, Hefei, Anhui 230026, China }
\centerline{$^{3}$Data Communication Science and Technology Research Institute, Beijing 100191, China}
\centerline{$^{4}$ Jinan Institute of Quantum technology, SAICT, Jinan 250101, China}
\centerline{$^{5}$ Shenzhen Institute for Quantum Science and Engineering, and Physics Department,}
\centerline{Southern University of Science and Technology, Shenzhen 518055, China}}
\begin{abstract}
\noindent 
We present improved results of twin-field quantum key distribution (TF-QKD) by using the structure of sending-or-not-sending (SNS) protocol and error rejection through two-way classical communications. Taking a typical experimental parameter setting, our method here improves the secure distance by 70 kilometers to more than 100 kilometers in comparison with the prior art results. Comparative study also shows advantageous in key rates at regime of long distance and large misalignment error rate for our method here. The numerical results show that our method here can have an advantageous key rates higher than various prior art results by 10 to 20 times. Taking all finite-key effects into consideration, it breaks the absolute repeater-less key rate bound with only $10^{11}$ pulses.
\end{abstract}

\maketitle
\section {Introduction}\label{def:sns}
\noindent Improving the secure key rate and the secure distance is the central issue of practical quantum key distribution (QKD)
~\cite{bennett2014quantum,shor2000simple,gisin2002quantum,kraus2005lower,gisin2007quantum,koashi2009simple,scarani2009security,duvsek2006quantum}
. In the recent years,  secure QKD in practice is extensively studied
~\cite{huttner1995quantum,brassard2000limitations,lu2000security,scarani2008quantum,pirandola2009direct,lydersen2010hacking,inamori2007unconditional,got2007sec,lu2002quantum}.
 In particular, the decoy-state
method ~\cite{hwang2003quantum,wang2005beating,lo2005decoy}
can beat the photon-number-splitting (PNS) attack~\cite{huttner1995quantum,brassard2000limitations,lu2000security} and guarantee the security with imperfect single-photon sources.
The method has been applied and studied extensively
~\cite{rosenberg2007long,schmitt2007experimental,peng2007experimental,liao2017satellite,peev2009secoqc,ad2007simple,wang2008experimental,wang2007quantum,dixon2010continuous,sasaki2011field,frohlich2013quantum,tamaki2014loss,xie2019optically,liu2019experimental1,tamaki2003unconditionally,xu2009experimental,hayashi2007upper,wang2008general,wang2009decoy,yu2016reexamination,boaron2018secure,chau2018decoy}.
Other protocols such as RRDPS protocol~\cite{sasaki2014practical,takesue2015experimental} were also proposed to beat PNS attack.  The detection loophole is closed by the measurement-device-independent (MDI)-QKD~\cite{braunstein2012side,lo2012measurement}.
In particular, with the decoy-state MDI-QKD \cite{
wang2013three,rubenok2013real,liu2013experimental,tang2014experimental,wang2015phase,comandar2016quantum,
yin2016measurement,wang2017measurement,curty2014finite,xu2013practical,xu2014protocol,
yu2015statistical,zhou2016making},  
we can make the protocol secure to beat the threats of both the imperfect single-photon sources and detection loopholes.

However, none of these protocols can exceed the linear scale of key rate \cite{takeoka2014fundamental,pirandola2017fundamental}, the fundamental limits such as the TGW bound \cite{takeoka2014fundamental} presented by Takeoka, Guha and Wilde, or the PLOB bound \cite{pirandola2017fundamental} established by Pirandola, Laurenza, Ottaviani, and Banchi.
The fascinating PLOB bound also makes the linear capacity of repeater-less key rate. 
In this work, we shall use PLOB bound for the criterion of repeater-less key rate.

Recently, Twin-Field Quantum-Key-Distribution(TF-QKD) protocol was proposed by Lucamarini et al. \cite{lu2018overcoming}. The key rate of this protocol $\sim O(\sqrt{\eta})$, where $\eta$ is the channel loss, thus its QKD distance has been greatly improved.  But security loopholes~\cite{wang2018effective,wang2018twin} are caused by the later announcement of the phase information in this protocol. Then many variants of TF-QKD ~\cite{wang2018twin,tamaki2018information,ma2018phase,lin2018simple,cui2019twin,curty2019simple,yu2019sending,lu2019twin,pirandola2019advances}
have been proposed to close those loopholes. To demonstrate those protocols, a series of experiments \cite{minder2019experimental,liu2019experimental,wang2019beating} have been done.
In particular, the sending-or-not-sending (SNS) protocol of TF-QKD given in Ref.~\cite{wang2018twin} has many advantages, such as the unconditional security, the robustness to the misalignment errors and so on. So far both the finite key effect has been fully studied \cite{yu2019sending,jiang2019unconditional}, also this protocol has been experimentally demonstrated in proof-of-principle in Ref.~\cite{minder2019experimental}, and realized in real optical fiber with the finite key effect being taken into consideration~\cite{liu2019experimental}. However, the protocol requests a small sending probability for both Alice and Bob and this limits its key rate and secure distance.  Here we report improved results for this SNS protocol using error rejection through randomly pairing and parity check.  We present improved method of SNS protocol based on its structure and the application of error rejection. Taking the finite key effect into consideration, we show that the method here can produce a key rate even higher than the  absolute limit of  key rate from repeater-less QKD with  whatever detection efficiency. We also make comparative study of different protocols numerically. It shows that our method here presents advantageous results at long distance regime and large noise regime, asymptotically or non-asymptotically. In particular, if taking the finite key effect into consideration with only $10^{11}$ total pulses, our method here can still break the absolute limit of repeater-less key rate with whatever  detection efficiency. Taking a typical experimental parameter setting, our method here improves the secure distance by 70 kilometers to more than 100 kilometers in comparison with the prior art results. Comparative study also shows advantageous in key rates at regime of long distance and large misalignment error rate for our method here. The numerical results show that our method here can have an advantageous key rates higher than various prior art results by 10 to 20 times. 

Before going into details, we first make a short review of the original SNS protocol, on some definitions and key rate formula.
 For conciseness, we shall use the italicized {\em they }  or {\em them} to represent Alice and Bob.
 Charlie takes the role of a quantum relay between Alice and Bob and he controls the measurement station.
\\{\bf 1, Different types of time windows .}
At each time, {\em they} each commit to a signal window with probability $p_Z$ or a decoy window with probability $p_X$ ($p_X+p_Z=1$).
When Alice (Bob) commits to a signal window, with probability $p$ she (he) decides {\em sending} and puts down a bit value 1 (0); with probability $(1-p)$ she (he)
decides {\em not-sending} and puts down a bit value 0 (1). Also, following  the decision of {\em sending} , she (he) sends out to Charlie a phase-randomized coherent state $|\sqrt{\mu_Z}\rangle \exp{(\mathbf{i}\theta'_m)}, m=1,2,3 \cdots$(In this paper, we denote the imaginary unit as $\mathbf{i}$, and $m$ corresponds to time window), where $\mu_Z$ is the intensity of a signal state, and $\theta'_m$ is the phase of the signal state (Alice and Bob choose their own $\theta'_m$ independently and privately, and the value of $\theta'_m$ in signal window will never be disclosed), and the phase can be different from time to time; following  the decision of {\em not-sending} , she (he) sends out a vacuum state, i.e., nothing.
When Alice (Bob) commits to a decoy window, she (he) sends out a decoy pulse in the phase-randomized coherent state randomly chosen from a few different intensities, $|{\sqrt{\mu_k}\exp{(\mathbf{i}\theta_{m})}}\rangle $,$k=1,2,3\cdots$ (Note that $\theta_m$ in decoy window will be publicly disclosed after the finish of whole communication).
\\{\em Remark 1.1} The phase-randomized coherent state is equivalent to a probabilistic mixture of different photon-number states.
When she  sends out a phase-randomized coherent state, it is possible that actually a vacuum or single-photon state or another photon-number state is sent out.
However, there is no confusion in the definition of bit value committed: it is the {\em decision} on {\em sending} or {\em not-sending} that determines the bit value. Once a decision is made, the bit value is determined  no matter what state is actually sent out.\\
{\em Remark 1.2} $Z$-window, $\tilde Z$-window, $\tilde Z_1$-window. A time window when both of {\em them} commit to the signal window is called a $Z$-window.  $\tilde Z$-windows are defined as the subset of $Z$-windows when one and only one party decides {\em sending}.
In SNS protocol, the random phases of the  coherent state of $Z$-windows sent out is never announced. As was mentioned already, the phase-randomized coherent state is equivalent to a classical mixture of different photon-number states. The $\tilde Z_1$-windows are a subset of $\tilde Z$-windows whenever a single-photon state  is actually sent out to Charlie.
\\
{\em Remark 1.3} $X$-window.  The time window is an $X$-window when {\em they} both commit to a decoy window and {\em } choose the same intensity for the decoy state. In an $X$-window, if both of {\em them} use the same intensity $\mu_k$, it is noted as an $X_k$-window. \\
{\bf 2, Effective events/effective windows.} The protocol requests Charlie to announce his measurement outcome. In particular, Charlie's announcement of one and only one detector clicking determines an {\em effective} time window or an {\em effective} event in the time window.\\
 For example, if an {\em effective} event happens in a certain $\tilde  Z_1$-window, the window is called an {\em effective} $\tilde Z_1$-window. If an {\em effective} event happens in a certain $X$-window, the window is called an {\em effective} $X$-window.
\\{\em Remark 2.1} Definition of bit-flip.  {\em They } will use bit values of {\em effective} $Z$-windows for final key distillation. In a $Z$-window when both of {\em them} make the decision of {\em sending}, or both of {\em them} make the decision of {\em not-sending}, {\em they} have actually committed to different bit values.
Therefore, an {\em effective} $Z$-window creates a bit-flip error when  both  of {\em them} make the decision of {\em sending}, or both  make the decision of {\em not-sending}. In particular, the bit-flip error rate $E$ of the SNS protocol is
\begin{equation}
E=\frac{n_{\mathbb{N}\mathbb{N}}+n_{\mathbb{SS}}}{n_t}
\end{equation}
and $n_{\mathbb{SS}},\; n_{\mathbb{NN}}$ are numbers of {\em effective} $Z$-windows when both sides decide {\em sending} and both sides decide {\em not-sending}, respectively. Value $n_t$ is the total number of {\em effective} $Z$-windows, and it can be directly observed. 
 Values of $n_{\mathbb{NN}}$, $n_{\mathbb{SS}}$ can be obtained by test on a few {\em effective} $Z$-windows as samples randomly taken from all
{\em effective} $Z$-windows. \\
{\bf 3, Un-tagged bits}
The bits from {\em effective} $\tilde Z_1$-windows are regarded as un-tagged bits. (More general case is given in Appendix.\ref{appendix:z0}). As was shown in Ref.~\cite{wang2018twin,yu2019sending,jiang2019unconditional},
 we can use the decoy-state analysis to verify faithfully $n_1$, the lower bound of the number of un-tagged bits, and  $\bar e_1^{ph}$, the upper bound of the phase-flip error rate ($e_1^{ph}$) of un-tagged bits. To make a tight estimation  of $\bar e_1^{ph}$, we need to post select only part of the {\em effective} $X_k$ windows by a certain phase-slice standard~\cite{hu2019sending}. With  these quantities, {\em they} can calculate the final  key length by formula
\begin{equation}\label{key0}
R=n_1 - n_1 H(\bar e_1^{ph}) -f n_t H(E),
\end{equation}
where $H(x)=-x\log_2x-(1-x)\log_2(1-x)$ is the binary Shannon entropy function, $f$ is the error correction coefficient which takes the value around $1.1$, and $n_t$ is the total number of {\em effective} $Z$-windows as defined earlier.

\section{ Bit-flip error rejection}\label{sec.3}

\subsection{ Refined structure of bit-flip error rate}\label{sec3.1}
As the bit-flip error rates of bits $0$ and bits $1$ are different in our protocol, we can refine the structure of bit-flip error rate to obtain higher key rate which is useful in the analysis of bit-flip error rejection, as shown below.

We label three types of {\em effective} events in $Z$-windows: 
\noindent 1. $C$ events are those events in {\em effective} $\tilde Z$-windows, and we use $n_C$ for the number of $C$ events;
among those $C$ events,
when Alice decides {\em sending} and Bob decides {\em not-sending}, Alice (Bob) obtains bits with value of 1 (1), and we define them as $C_1$ events,  and we denote the total number of those events by $n_{C_1}$;
among those $C$ events,
when Alice decides {\em not-sending} and Bob decides {\em sending}, Alice (Bob) obtains bits with value of 0 (0), we define them as $C_0$ events,  and we denote the total number of those events by $n_{C_0}$; note that $n_{C_1}+n_{C_0}=n_C$;
\noindent 2. $D$ events, both parties decide {\em sending}, and the total number of $D$ events is $n_D$;
\noindent 3. $V$ events, the remaining {\em effective} events, the total number of $V$ events is $n_V$.
Though we divide the {\em effective} events in $Z$-windows into three sets here, Alice and Bob cannot tell which set each {\em effective} event is from. {\em They} can only know some features about these sets.
Clearly, the bit-flip error rates of bits $0$ and bits $1$ are different.

Bob divides his bits in $Z_B$ into two groups: group of bits $0$ and group of bits $1$.
The group of bits $0$ contains $N_{0}=n_D+n_{C_0}$ bits, and its bit-flip error rate is $E_{0}=n_D/N_{0}$; the group of bits $1$ contains $N_{1}=n_V+n_{C_1}$ bits, and its bit-flip error rate is $E_{1}=n_V/N_{1}$.
Then we have the following improved key length formula
\begin{equation}\label{equ:keyrate1}
	N_f = n_{1}-n_1H(\bar e_1^{ph})-f[N_{0}H(E_{0})+N_{1}H(E_{1})],
\end{equation}
where the definition of $n_1$ and $\bar e_1^{ph}$ are the same as Eq.(\ref{key0}). Note that $N_0,E_0,N_1$ and $E_1$ can be observed directly.
Although the new key-length formula based on refined data structure of SNS can somehow improve the key rate, it does not really make a significant improvement
 unless we use error rejection through randomly pairing and parity checks with two-way classical communications\cite{gottesman2003proof,chau2002practical,wang2004quantum} as shown below.

\subsection{bit-flip error rejection}\label{sec3A}

After Alice (Bob) gets the string $Z_A$ ($Z_B$) of length $n_t$ $(n_t=n_C+n_D+n_V)$, {\em they} make randomly pairing: Bob randomly groups his bits two by two and announces his grouping information to Alice. Alice takes the same random grouping accordingly. Then {\em they} have $\lfloor\frac{n_t}{2}\rfloor$ bit pairs. {\em They} then compare the  parity of pairs. Although there is no bit-flip error for those un-tagged bits, there are bit-flip errors of those tagged bits. Bit-flip Error rejection (BFER): After the parity check, {\em they} give up the whole pair if {\em they} find  different parity values of two sides and {\em they} keep the first bit and discard the second bit if {\em they} find the same parity values of two sides. The bits kept from the events of same parity values in the parity check are named as the {\em survived bits}.

Instead of using the averaged bit-flip error rate, we shall use the refined bit-flip error rate. Before parity check operation, Bob can classify his bit pairs after random pairing. If both bits in a pair are from $C$ events, we label this pair as a $CC$-pair, and the total number of this kind of pairs is $n_{CC}$. Similarly, if the first bit of a pair is from $\alpha$ and the second bit of a pair is from $\beta$, we label the pair by $\alpha \beta$, and the number of this kind of pairs are denoted by $n_{\alpha \beta}$. There are 9 possible different labels, say $CC, VV, DD, CV, VC, CD, DC, VD, DV$.

After BFER, Alice and Bob use the survived bits to form new strings $\hat{Z}_A$ and $\hat{Z}_B$. The number of survived bits is:
\begin{equation}\label{eq:random}
	\tilde{n}_t=n_{CC}+n_{VD}+n_{DV}+n_{VV}+n_{DD},
\end{equation}
where $n_{\alpha\beta}$ is the number of $\alpha\beta$-pairs ($\alpha\beta=CC,VD,DV,VV,DD$). Since $C$ events can be further subdivided into $C_0$ and $C_1$ as described in Sec.~\ref{sec3.1}, $CC$-pairs can be further subdivided into $C_1C_1, C_1C_0, C_0C_1$ and $C_0C_0$, the corresponding number of them have relationship: $n_{CC}=n_{C_1C_1}+n_{C_1C_0}+n_{C_0C_1}+n_{C_0C_0}$.

Bob divides the bits in $\hat{Z}_B$ into three classes: bits in class 1 are originally from odd-parity pairs. Denote $n_{t1}$ as the number of bits in this class, and $E_1$ as the corresponding bit-flip error rate.

Bits in class 2 are originally from even-parity pairs, and both bits in a pair are all $0$. Denote $n_{t2}$ as the number of bits in this class, and $E_2$ as the corresponding bit-flip error rate.
Class 3 contains the remaining bits. Denote $n_{t3}$ as the number of bits in this class, and $E_3$ as the corresponding bit-flip error rate.

As the major consequence of the error rejection\cite{gottesman2003proof,chau2002practical,wang2004quantum}, the bit-flip error rate is expected to be reduced. To see how much it is reduced, we can use the following iteration formula for the expected bit-flip error rate after parity check.  Note that, {\em they} will only keep the first bit if the pair passes the parity check (same parity
values at two sides) and {\em they} discard the whole parity pair if it fails to pass the parity check (different parity values at two sides). As we can see, before error rejection, the bit flip error rate is $E_Z=(n_D+n_V)/n_t$.
After this error rejection, the number of bits from each classes and the corresponding error rates are expected by:
\begin{equation}\label{equ:split}
 	\begin{split}
 	&n_{t1}=n_{VD}+n_{DV}+n_{C_1C_0}+n_{C_0C_1}, E_1=\frac{n_{VD}+n_{DV}}{n_{t1}},\\
 	&n_{t2}=n_{DD}+n_{C_0C_0},   E_2=\frac{n_{DD}}{n_{t2}},\\
 	&n_{t3}=n_{VV}+n_{C_1C_1}, 	E_3=\frac{n_{VV}}{n_{t3}}. \\
 	\end{split}
 \end{equation}
 
\subsection{Information leakage, phase-flip error rate, and secure key rate}
Although the parity check above can reduce the bit-flip error rate, it also makes the information leakage to the remaining bits. To make a secure final key, we have to consider this consequence. For this goal, we need consider the new phase-flip error rate of those un-tagged bits after the error rejection.
  Prior to the parity check, if both bits in a pair are un-tagged bits,
 then the remaining bit after bit-flip error rejection process is an un-tagged bit. Note that, since an un-tagged bit prior to the error rejection can be only from an {\em effective} $\tilde Z_1$-window, there is no bit-flip error and those pairs containing two un-tagged bits will be for sure to pass the parity check.  Using the tagged model\cite{inamori2007unconditional,got2007sec}, we don't have to know which remaining bits are un-tagged. We only need to know how many of them are un-tagged. Also, it doesn't matter for security if we underestimate the number of un-tagged bits or overestimate the number of tagged bits.
 For simplicity,  we shall regard a survived bit as an un-tagged bit by this criterion that prior to the parity check both bits in the same group are un-tagged bits. We shall regard all other survived bits as tagged bits.  After BFER with parity check, the number of un-tagged bits in string $\hat{Z}_B$ is
\begin{equation}\label{untaggedbits}
\tilde{n}_1=\frac{n_t}{2}(\frac{n_1}{n_t})^2.
\end{equation}
Since there are backward action\cite{gottesman2003proof}, the phase-flip error rate for the remaining bits after error rejection changes\cite{chau2002practical}.
The phase-flip error rate for the bits of type I in Ref.\cite{chau2002practical} can be iterated by the standard formula\cite{chau2002practical} with the specific setting of bit-flip error rate being 0 here:
\begin{equation}\label{phase}
\tilde{e}_1^{ph}=2\bar e_1^{ph}(1-\bar e_1^{ph}).
\end{equation}
Actually, we can also directly obtain the iterated formula above for phase-flip error rate through a virtual protocol with quantum entanglement.
The key length formula after error rejection is
\begin{equation}\label{key:random}
N_f=\tilde{n}_1 [1-H(\tilde{e}_1^{ph})]-f[n_{t1}H(E_1)+n_{t2}H(E_2)+n_{t3}H(E_3)].
\end{equation}

\subsection{key rate with finite key effects}
If we take the finite key effects into consideration, using the method given in \cite{curty2014finite,jiang2019unconditional},the key length formula after error rejection can be expressed as
\begin{equation}\label{key_random}
\begin{split}
N_f&=n_{1L} [1-H(e_{1u}^{ph})]-f[n_{t1}H(E_1)+n_{t2}H(E_2)\\
+&n_{t3}H(E_3)]-\log_2\frac{2}{\varepsilon_{cor}}-2\log_2\frac{1}{\sqrt{2}\varepsilon_{PA}\hat{\varepsilon}},
\end{split}
\end{equation}
where the definition of $n_{t1},n_{t2},n_{t3}$ and $E_1,E_2,E3$ are all given in Sec.\ref{sec3A}.  They are all directly observable, though one can theoretically forcast them by Eq.(\ref{equ:split}).
With Eq.(\ref{key_random}),
the protocol is $\varepsilon_{tot}$-secure with $\varepsilon_{tot}=\varepsilon_{cor}+\varepsilon_{sec}$,
where $\varepsilon_{sec}=2\hat{\varepsilon}+4\overline{\varepsilon}+\varepsilon_{PA}+\varepsilon_{n1}$, and $\varepsilon_{cor}$ is the failure probability of error correction. Here $\hat \varepsilon$ is the coefficient when using the chain rules for smooth min- and max- entropies\cite{vitanov2013chain}, $\varepsilon_{PA}$ is the failure probability of privacy amplification,  $\overline{\varepsilon}$ is the failure probability for estimation of the phase-flip error rate of un-tagged bits in string  $\hat{Z}_A$ and  $\hat{Z}_B$, and $\varepsilon_{n1}$ is the failure probability for estimation of the lower bound of the total number of un-tagged bits in string $\hat{Z}_A$ and  $\hat{Z}_B$. In this paper, we set the failure probability of Chernoff Bond \cite{chernoff1952measure,curty2014finite}  as $\xi=10^{-10}$, and $\overline{\varepsilon}=3\xi$,$\varepsilon_{n_1}=6\xi$. And we set $\varepsilon_{cor}=\hat{\varepsilon}=\varepsilon_{PA}=\xi$, thus $\varepsilon_{tot}=2.2\times 10^{-9}$.

In Eq.(\ref{key_random}), $n_{1L}$ is lower bound of the number of un-tagged bits after error rejection, and $e_{1u}^{ph}$ is upper bound of phase-flip error rate of un-tagged bits after error rejection, explicitly they can be calculated by
\begin{equation}
\begin{split}
n_{1L}=&\varphi^L(\langle{\tilde{n}_1}\rangle)\\
\tilde{e}^{ph}_{1u}=&\frac{\varphi^U(\langle{\tilde{n}_1}\rangle2\langle{\bar e_1^{ph}}\rangle(1-\langle{\bar e_1^{ph}}\rangle)}{\langle{\tilde{n}_1}\rangle},
\end{split}
\end{equation}
where
\begin{equation}
\langle{\tilde{n}_1}\rangle=\frac{n_t}{2}(\frac{\langle{n_1}\rangle}{n_t})^2,
\end{equation}
and the definition of $n_t$ is the same as Eq.(\ref{key0}). Here $\langle{n_1}\rangle$ is the expected value of the number of un-tagged bits before error-rejection, and $\langle{\bar e_1^{ph}}\rangle$ is the expected value of the upper bound of phase-flip error rate of un-tagged bits before error-rejection. $\langle{n_1}\rangle$ and $\langle{\bar e_1^{ph}}\rangle$ can be calculated through four intensities decoy state method, as given in Ref.\cite{yu2019sending,jiang2019unconditional}. And we use Chernoff-bound \cite{chernoff1952measure,curty2014finite} $\varphi^{L}(Y)$ to calculate the lower bound of value $Y$, and $\varphi^{U}(Y)$ to calculate the Upper bound of value $Y$. In specific
\begin{equation}
\begin{split}
\varphi^U(Y)=[1+\delta_1(Y)]Y\\
\varphi^L(Y)=[1-\delta_2(Y)]Y
\end{split}
\end{equation}
where $\delta_1$ and $\delta_2$ are obtained by solving the following equations
\begin{equation}
\begin{split}
(\frac{\exp{\delta_1}}{(1+\delta_1)^{1+\delta_1}})^{Y}=\frac{\xi}{2}\\
(\frac{\exp{\delta_2}}{(1-\delta_2)^{1-\delta_2}})^{Y}=\frac{\xi}{2},\\	
\end{split}
\end{equation}
where $\xi$ is the failure probability of Chernoff-bound.

\subsection{numerical simulation}
\begin{figure}\centering
	\includegraphics[width=285pt]{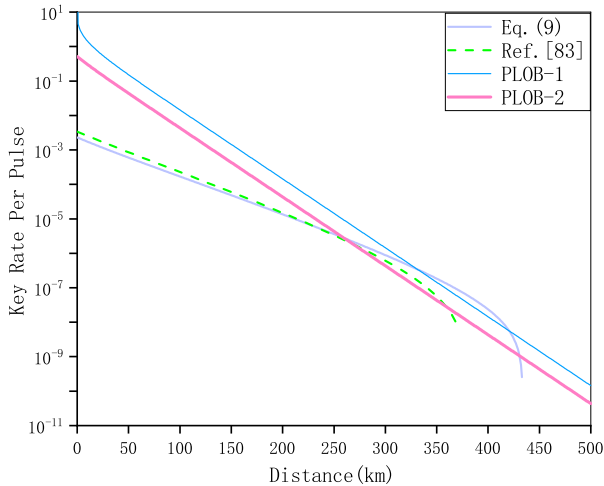}
	\caption{(Color online)Optimized key rates with $10^{11}$ pulses by Eq.(\ref{key_random}) (purple solid line)
		and the method of Ref.~\cite{maeda2019repeaterless} (green dash line). PLOB-1 is repeter-less key rate bound \cite{pirandola2017fundamental} with detector efficiency $\eta=1$, i.e., the absolute limit of repeater-less key rate. PLOB-2 is repeter-less key rate bound \cite{pirandola2017fundamental} with detector efficiency $\eta=0.3$, i.e., the relative limit of repeater-less key rate.
		Here, four intensities\cite{yu2019sending,jiang2019unconditional} are used in the decoy state calculation with Eq.(\ref{key_random}). 
		Device parameters are given by row A of Table \ref{tab:parameters}. Finite key effects are taken with failure probability of $\xi=10^{-10}$.}\label{fig:1e1}
\end{figure}

\begin{figure}\centering
	\includegraphics[width=285pt]{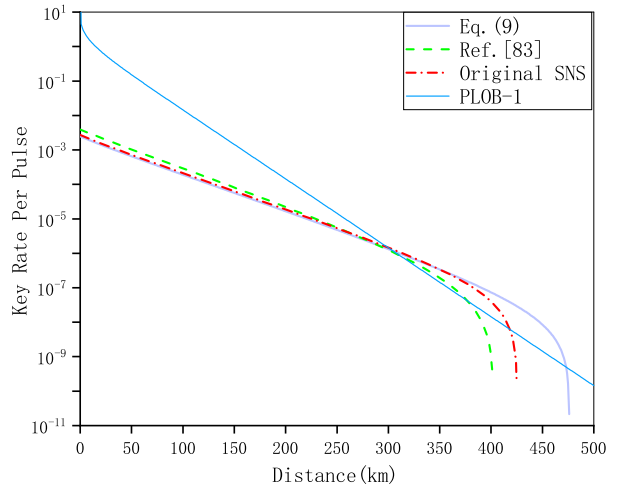}
	\caption{(Color online)Optimized key rates with $10^{12}$ pulses by Eq.(\ref{key_random}) (purple solid line),
		original SNS protocol\cite{wang2018twin} with finite key effects\cite{jiang2019unconditional}(red dots dash line)
		and  the method of Ref.~\cite{maeda2019repeaterless} (green dash line). PLOB-1 is repeter-less key rate bound \cite{pirandola2017fundamental} with detector efficiency $\eta=1$, i.e., the absolute limit of repeater-less key rate. 
		Here, four intensities\cite{yu2019sending,jiang2019unconditional} are used in the decoy state calculation with Eq.(\ref{key_random}) and the original SNS protocol\cite{wang2018twin}. 
		Device parameters are given by row B of Table \ref{tab:parameters}.  Finite key effects are taken with failure probability of $\xi=10^{-10}$.}\label{fig:1e2}
\end{figure}

Here in our simulation, we set the failure probability for parameters estimation as $\xi=10^{-10}$ for finite key effects in the key-rate calculation of Fig.\ref{fig:1e1} and Fig.\ref{fig:1e2}.

As we can see in Fig.\ref{fig:1e1}, when the finite key effect is taken into consideration, our improved method given in Eq.(\ref{key_random}) presents the advantageous result at long distance regime. It exceeds the absolute limit of repeater-less key rate (PLOB bound\cite{pirandola2017fundamental} with detector efficiency $\eta=1$), also the secure distance is 70km longer compared with Ref.\cite{maeda2019repeaterless}. At shorter distance regime, Ref.\cite{maeda2019repeaterless} produces a higher key rate. Device parameters used are given in row A of Table.\ref{tab:parameters}

In Fig.\ref{fig:1e2}, setting the total number of pulses as $10^{12}$, our improved method given in Eq.(\ref{key_random})  breaks the absolute limit of repeater-less key rate more significantly. Eq.(\ref{key_random}) gives the longest secure distance, 50km longer than Original SNS protocol\cite{wang2018twin} and 70km longer than Ref.\cite{maeda2019repeaterless} under the same parameters given in row B of Table.\ref{tab:parameters}. 


\begin{table}[htb]
	\begin{ruledtabular}
		\begin{tabular}{|l|l|l|l|l|l|}
			& $d$          & $\eta_0$     &$f$        &$e_a$   &$N$\\ \hline
			A &$10^{-8}$      &$0.30$      &$1.10$     &$0.03$ &$10^{11}$\\ \hline
			B &$10^{-8}$      &$0.30$      &$1.10$     &$0.03$ &$10^{12}$\\ \hline
			C&$10^{-8}$      &$0.50$      &$1.15$      &$0.05$ &$asymptotic$\\ \hline
			D &$8\times10^{-8}$&$0.30$     &$1.15$     &$0.05$ &$asymptotic$\\ \hline
			E &$8\times10^{-8}$&$0.30$     &$1.15$     &$0.10$ &$asymptotic$\\ \hline
			F &$8\times10^{-8}$&$0.30$      &$1.15$     &$0.15$ &$asymptotic$\\ \hline
		\end{tabular}
	\end{ruledtabular}
	\caption{Device parameters used in numerical simulations.
		$d$: the dark count rate.
		$\eta_0$: the detection efficiency of all detectors.
		$f$: the error correction inefficiency.
		$e_a$: the misalignment error.
	    $N$: total number of time windows.
	    The fiber loss is setted as $\alpha_f=0.2$.
    }\label{tab:parameters}.
\end{table}

\begin{table}[htb]
	\begin{ruledtabular}
		\begin{tabular}{|c|c|c|}
			distance(km)  &Ref.\cite{fang2019surpassing}         &Eq.(\ref{key_random}) \\
			\hline
			502           &$1.68\times 10^{-9}$      &$1.86\times10^{-8}$ \\ 
		\end{tabular}
	\end{ruledtabular}
	\caption{The key rates of Ref.\cite{fang2019surpassing} and Eq.(\ref{key_random}) in 502 km.  We use the parameters of Ref.\cite{fang2019surpassing} in calculation, e.g., the dark count rate is $d=1.26\times 10^{-8}$, the misalignment-error probability is $e_d=9.8\%$, the detection efficiency is $\eta_0=0.29$, the fiber loss is $\alpha_f=0.162$, the failure probability is $\xi=1.71\times 10^{-10}$, and the total number of pulses is $N=2.0\times10^{13}$.
	}\label{tab:502km}
\end{table}

In Table.\ref{tab:502km}, the key rates of Eq.(\ref{key_random}) and Ref.\cite{fang2019surpassing} in 502km are given. We use the parameters of Ref.\cite{fang2019surpassing} in calculation, which are $d=1.26\times 10^{-8}$, $e_d=9.8\%$, $\eta_0=0.29$, $\alpha_f=0.162$, $\xi=1.71\times 10^{-10}$ and $N=2.0\times10^{13}$. Table.\ref{tab:502km} shows that the key rate of Eq.(\ref{key_random}) is $10$ times higher than that of Ref.\cite{fang2019surpassing} in 502km.

\section{odd-parity error rejection and AOPP method}

\begin{table}[htb]
	\begin{ruledtabular}
		\begin{tabular}{|c|c|c|c|c|}	
			distance(km)  &$E_Z$         &$E_1$                   &$E_2$        &$E_3$\\
			\hline
			100           &$10.28\%$      &$2.27\times10^{-6}$     &$5.00\%$     &$9.77\times10^{-11}$\\
			\hline
			300           &$10.28\%$	 &$2.33\times10^{-4}$     &$4.96\%$     &$1.04\times10^{-6}$\\
			\hline
			500           &$13.00\%$     &$2.03\%$                &$3.48\%$     &$1.18\%$  \\
		\end{tabular}
	\end{ruledtabular}
	\caption{Comparison of the bit-flip error rate prior$(E_Z)$(defined in Sec.\ref{sec3A}) and after bit-flip error rejection $(E_1,E_2,E_3)$ under typical distances.
		Device parameters are given in row C of Table \ref{tab:parameters}.}\label{tab:example2}
\end{table}

In Table.~\ref{tab:example2}, we list the bit-flip error rate before and after BFER in some typical distances, the device parameters are given in row C of Table \ref{tab:parameters}. We can see that, with the help of bit-flip error rejection and the refined structure of bit-flip error rate, the bit-flip error rate is reduced dramatically, hence our method can significantly improve the performance of the method in Ref.~\cite{wang2018twin}.

In the original SNS protocol, we have to use very small sending probability so as to control the bit-flip error rate. By using BFER here, we can improve the sending probability and therefore obtain advantageous result at the regime of long distances. But, consider Eq.(\ref{key:random}), after error rejection, the bit-flip error rate of survived bits from even-parity  groups can be still quite large. This limits the key rate. Naturally, more advantageous results can be obtained if we use odd-parity events only.

\subsection{odd-parity sifting}\label{odd_sfiting}
 Note that after post-selection process described above, the bit-flip error rate is concentrated on those  even-parity pairs, as shown in Table.~\ref{tab:example2}.  Thus if we only use those odd-parity pairs   to extract the final keys, the final key rates may be improved. Based on this idea, we continue the operation of Sec.~\ref{sec.3}, but in the final step, only the odd-parity pairs are reserved to extract the final keys. After random pairing, the total number of odd-parity pairs $N_R$ would be
\begin{equation}\label{eq:randompairing}
N_R=\frac{N_1N_0}{N_1+N_0},
\end{equation}
where the definitions of $N_1$ and $N_0$ are the same as Eq.(\ref{equ:keyrate1}). Note that $N_{R}$ is directly observable.

After the bit-flip error rejection and odd-parity sifting, Alice and Bob use the remaining bits to form new strings $\hat{Z}'_A$ and $\hat{Z}'_B$.

We denote the total number of un-tagged bits in $C_1$ events by $n_1^1$,  and the total number of un-tagged bits in $C_0$ events by $n_1^0$, then we have $n_1=n_1^1+n_1^0$.  The number of un-tagged bits $n_1^0, n_1^1$ can be estimated exactly in the asymptotic case.

After the bit-flip error rejection and odd-parity sifting, the number of un-tagged bits in string $\hat{Z}'_B$ is
\begin{equation}\label{eq:untagged1}
{n'}_1=\frac{n_t}{2}[\frac{n_1^1n_1^0}{n_t^2}+\frac{n_1^0n_1^1}{n_t^2}].
\end{equation}
with the non-trivial proof in the Appendix.~\ref{appendix:odd}, we have the following iteration formula for the phase-flip error rate of the survived bits after error rejection taken from those odd-parity groups only:
\begin{equation}\label{phase_odd}
	e'^{ph}_{odd}=2\bar e^{ph}_1(1-\bar e^{ph}_1).
\end{equation}
And hence the key rate formula
\begin{equation}\label{key:odd}
N_f={n'}_1[1-H(e'^{ph}_{odd})]-fn_{t1}H(E_{1}),
\end{equation}
where the definition of $n_{t1}$ and $E_1$ are the same as Eq.(\ref{key:random}), and they are all directly observable. The phase error rate iteration formula Eq.(\ref{phase_odd}) happens to be the same with Eq.(\ref{phase}). However, the proof of this is nontrivial.

Note that, after error rejection, if we only use those survived bits from odd-parity groups, the original phase-flip iteration formula Eq.(\ref{phase}) does not have to hold automatically, because it is for the case of using both odd-parity events and  even-parity events, while the phase-flip error rate for survived bits from odd-parity group can be different from  that of even-parity group.
Consider the specific example: Alice and Bob initially share a number of entangled pairs with each of them being in the identical state $|{\psi}\rangle=\frac{1}{\sqrt{2}}(|{00}\rangle+\exp({\mathbf{i}\phi})|{11}\rangle)$, and $\sigma=|{\psi}\rangle\langle{\psi}|$,which is
\begin{equation}
	\begin{split}
	\sigma=&\frac{1}{2}[|{00}\rangle\langle{00}|+\exp({\mathbf{i}\phi})|{11}\rangle\langle{00}|\\
	+&\exp({-\mathbf{i}\phi})|{00}\rangle\langle{11}|+|{11}\rangle\langle{11}|].
	\end{split}
\end{equation}
One can easily check that after pairing and parity check, the phase-flip error rate for survived pairs from odd-parity groups is 0, while the value from even-parity groups is $2\sin ^2\phi$. They are {\em different}.
Therefore,
in general, we need a separate proof for the iteration formula of phase-flip error rate of survived bits from odd-parity groups only. We complete this non-trivial proof in Appendix\ref{appendix:odd}

\subsection{Actively Odd-Parity Pairing}\label{sec.AOPP}

 Further, if we actively make odd-parity pairing, say, in our randomly grouping, we let each group contain a pair of different bits, we shall obtain more odd-parity pairs than passively choose odd-parity events after randomly pairing.  First, we define ``actively odd-parity pairing'' (AOPP) process: Randomly group the bits from a certain bit string two by two, with a condition that each pairs are for sure in odd-parity. That is to say, the random grouping here is not entirely random. If one chooses the first bit for a certain pair entirely randomly from all available bits, then the second bit can only be chosen randomly from those available bits which have a bit value {\em different} from the first bit.  Obviously, {\em they} can repeat the above AOPP procedure until {\em they} obtain the largest possible number of odd-parity bit pairs.

After Bob gets the string $Z_B$, he performs AOPP.  Then the total number of odd-parity pairs he can obtain through AOPP is
\begin{equation}\label{eq:activepairing}
\begin{split}
N_A=\min{(N_1,N_0)},
\end{split}
\end{equation}
where the definitions of $N_1 ,N_0$ are the same as Eq.(\ref{equ:keyrate1}).
Note that $N_1,N_0,N_A$ can be observed directly. After parity check, there are two possible different groups, labeled with $C_1C_0+C_0C_1,VD+DV$. The total number pairs in each group
\begin{equation}
\begin{split}
&N_{C_1C_0+C_0C_1}=\frac{n_{C_1}}{N_1}\frac{n_{C_0}}{N_0}N_A,\\
&N_{VD+DV}=\frac{n_V}{N_1}\frac{n_D}{N_0}N_A.
\end{split}
\end{equation}
{\em They} keep the first bit in the pair, and discard the other bit.
Then {\em they} use the remaining bits to form new string $\hat{Z}''_A$ and $\hat{Z}''_B$ with length
\begin{equation}\label{eq:active}
\begin{split}
\tilde{N}_t=N_{C_1C_0+C_0C_1}+N_{VD+DV},
\end{split}
\end{equation}
and the bit-flip error rate
\begin{equation}
{E''}_Z=\frac{N_{VD+DV}}{\tilde{N}_t}.
\end{equation}
After AOPP and parity check, the number of un-tagged bits in string $\hat{Z}''_B$ is
\begin{equation}
\begin{split}
{n''}_{1}=\frac{n_1^0}{N_0}\frac{n_1^1}{N_1}N_A.
\end{split}
\end{equation}
As proved in Appendix.\ref{appendix:AOPP}, the phase-flip error rate should be the same as Eq.(\ref{phase_odd}), which is
\begin{equation}\label{phase:AOPP}
{e''}_{1}^{ph}=2\bar e_1^{ph}(1-\bar e_1^{ph}),
\end{equation}
and the definitions of $n_1^0$ and $n_1^1$ are the same as Eq.(\ref{eq:untagged1}), and the asymptotic key length formula
\begin{equation}\label{key:AOPP}
N_f={n''}_{1}[1-H({e''}_{1}^{ph})]-f \tilde{N}_t H({E''}_Z).
\end{equation}

\begin{figure}\centering
	\includegraphics[width=285pt]{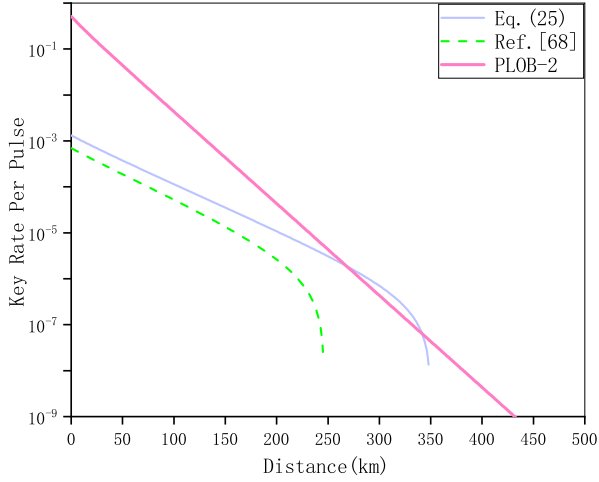}
	\caption{(Color online) The optimized asymptotic key tates by Eq.(\ref{key:AOPP}) (the purple solid line) and NPPTF-QKD\cite{cui2019twin} (green dash line). PLOB-2 is repeter-less key rate bound \cite{pirandola2017fundamental} with detector efficiency $\eta=0.3$, i.e., the relative limit of repeater-less key rate.
 	The method of this work presents advantageous results in long distance regime.
	Devices' parameters are given by row F of Table \ref{tab:parameters}. }\label{fig:1e3}
\end{figure}

\begin{figure}\centering
	\includegraphics[width=285pt]{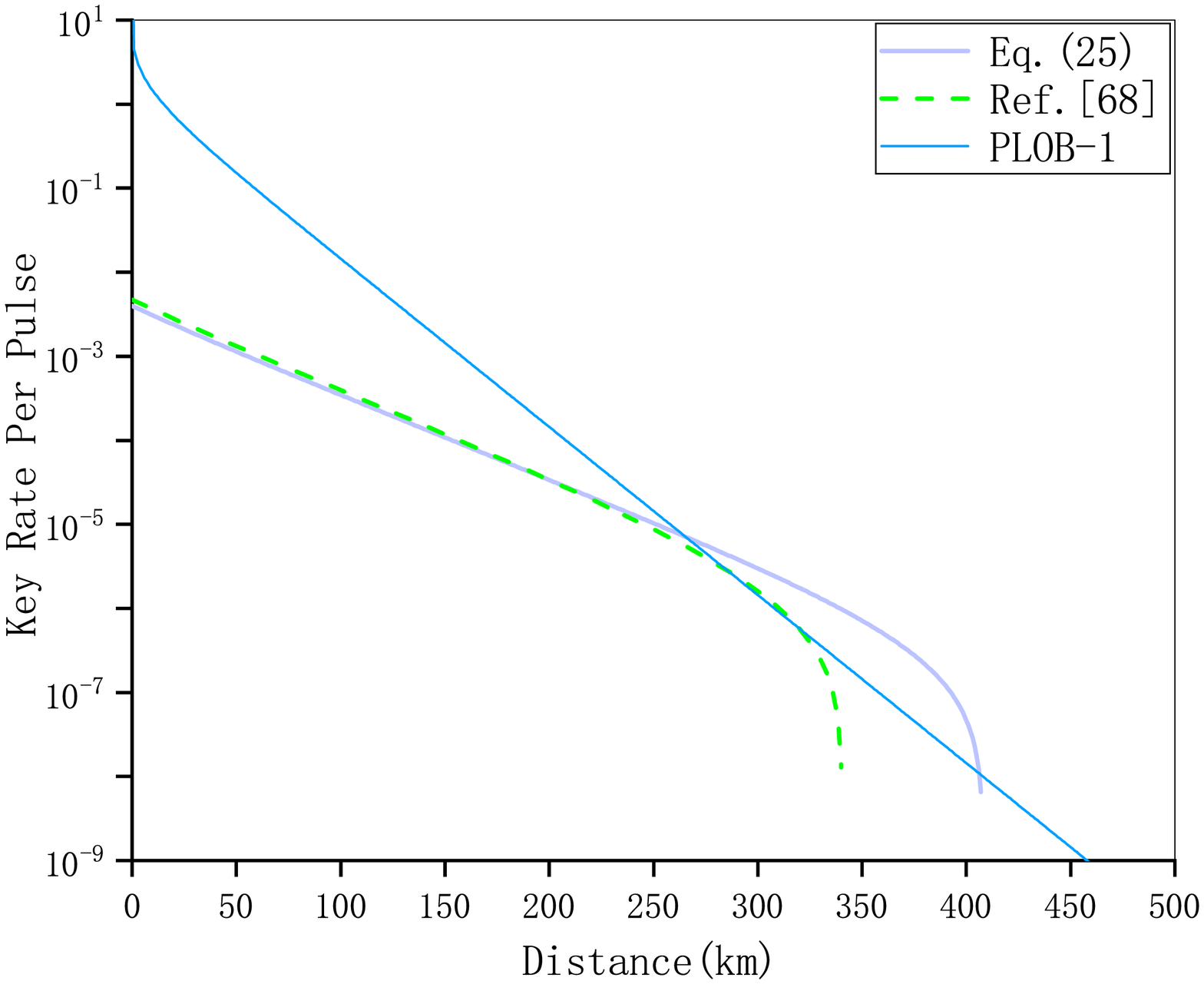}
	\caption{(Color online)  The optimized asymptotic key tates by Eq.(\ref{key:AOPP}) (the purple solid line) and NPPTF-QKD\cite{cui2019twin} (green dash line). PLOB-1 is repeter-less key rate bound \cite{pirandola2017fundamental} with detector efficiency $\eta=1$, i.e., the absolute limit of repeater-less key rate. 
		The method of this work presents advantageous results in long distance regime.
		Devices' parameters are given by row D of Table \ref{tab:parameters}. }\label{fig:1e4}
\end{figure}

\begin{figure}\centering
	\includegraphics[width=285pt]{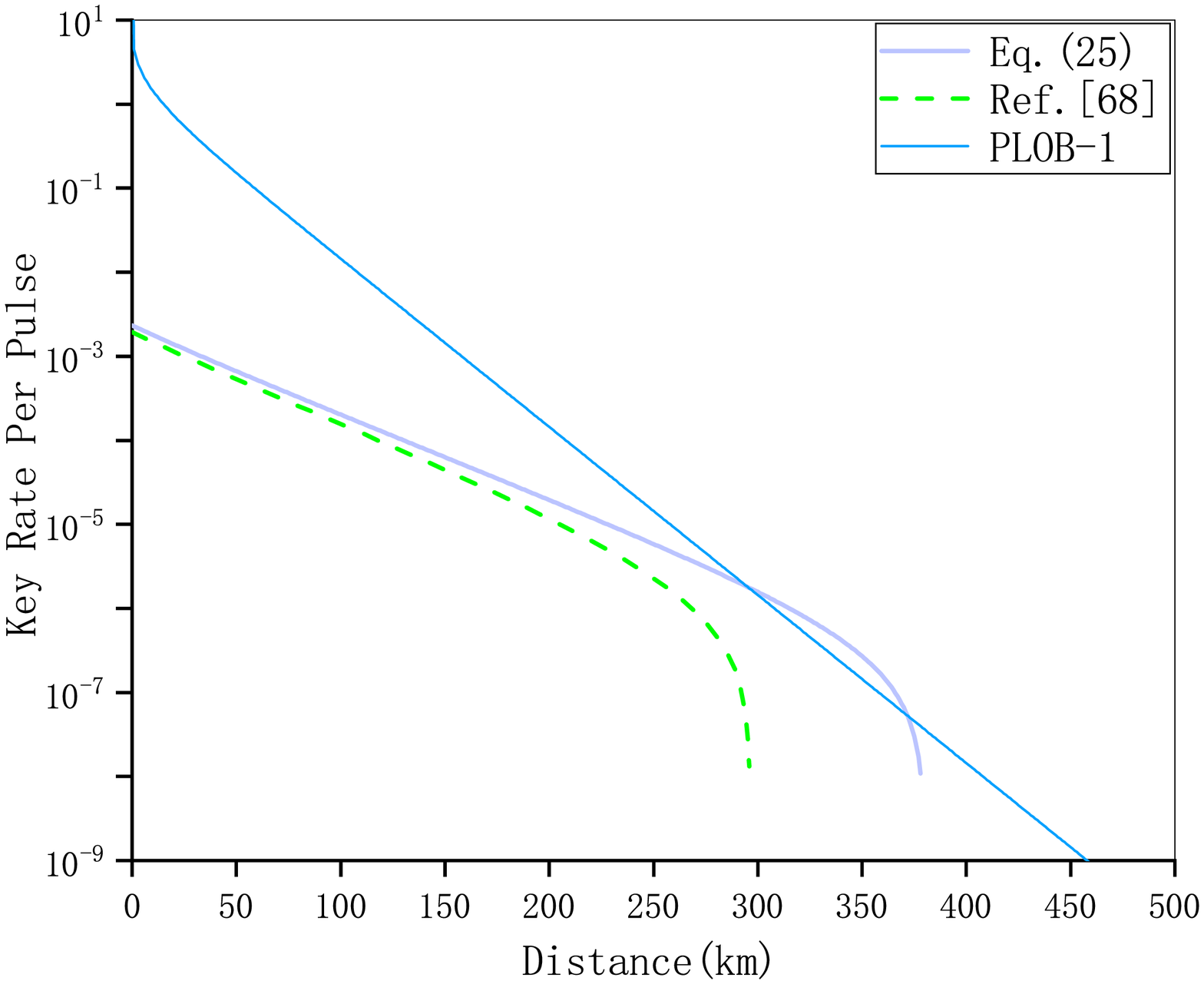}
	\caption{(Color online)  The optimized asymptotic key tates by Eq.(\ref{key:AOPP}) (the purple solid line) and NPPTF-QKD\cite{cui2019twin}(green dash line). PLOB-1 is repeter-less key rate bound \cite{pirandola2017fundamental} with detector efficiency $\eta=1$, i.e., the absolute limit of repeater-less key rate.
	The method of this work presents advantageous results in long distance regime.
	Devices' parameters are given by row E of Table \ref{tab:parameters}. }\label{fig:1e5}
\end{figure}

\begin{table}[htb]
	\begin{ruledtabular}
		\begin{tabular}{|c|c|c|c|}
			distance(km)& $160$     &$240$                  &$300$\\
			\hline
			Eq.(\ref{key:AOPP})    &$2.79\times10^{-5}$ &$3.99\times10^{-6}$  &$7.01\times10^{-7}$\\
			\hline
			Ref.~\cite{cui2019twin}  &$1.00\times10^{-5}$ &$1.67\times10^{-7}$  &-\\
		\end{tabular}
	\end{ruledtabular}
	\caption{ The optimized asymptotic key rates by Eq.(\ref{key:AOPP}) and NPPTF-QKD\cite{cui2019twin} in some typical distance.
	Device parameters are given in row F of Table \ref{tab:parameters}.}\label{tab:NPPTF}
\end{table}

In the following asymptotic key-rate calculations, we assume infinite intensities for the decoy-state analysis.
In Fig.~\ref{fig:1e3}, we can see that when the misalignment is as large as $15\%$, our improved SNS protocol can still break the relative limit of repeater-less key rate\cite{pirandola2017fundamental}, and the total length of secure distance given by Eq.(\ref{key:AOPP})is 100km further than that given by Ref.\cite{cui2019twin}. Besides, from Table.\ref{tab:NPPTF} we can see that the key rates of Eq.(\ref{key:AOPP}) is about 20 times higher than that in Ref.\cite{cui2019twin} in 240km.

In Fig.~\ref{fig:1e4}, we can see that when the dark count is as large as $8\times10^{-8}$, and the detection efficiency is $0.3$, our improved post data processing method, the AOPP method, Eq.(\ref{key:AOPP}) can make SNS protocol  break the absolute limit of repeater-less key rate\cite{pirandola2017fundamental}, and the key rate is higher than that of NPPTF-QKD\cite{cui2019twin} at long distance regime.   Device parameters are given in row D of Table.~\ref{tab:parameters}.  

Our result in this work has even more advantageous results in the regime of larger misalignment error rate. This property can be rather useful in situation such as field test, the free space realization, and so on.  In Fig.~\ref{fig:1e5}, we further increase the optical misalignment error to $10\%$, key rates of AOPP Eq.(\ref{key:AOPP}) can still break the absolute limit of repeater-less key rate\cite{pirandola2017fundamental}, and the key rate is higher than that of NPPTF-QKD\cite{cui2019twin} at large noise regime. Device parameters are given in row E of Table.~\ref{tab:parameters}.

\section{conclusion and discussion}\label{sec.5}
\noindent In this paper, we have improved the performance of the SNS protocol \cite{wang2018twin} with the help of the post processing of bit-flip error rejection with two-way classical communications method \cite{chau2002practical,gottesman2003proof,wang2004quantum}. We have considered finite key effect in Sec.\ref{sec.3}, and the simulation results show that our methods in this work give advantageous results at long distance regime: with the finite key effects being taken, even the dark counting rate is $10^{-8}$, the key rate of this work can still exceed the absolute limit of repeater-less key rate\cite{pirandola2017fundamental}, and the secure distance is improved 70km compared with PM-QKD\cite{maeda2019repeaterless}. To achieve better performance at all distance points, we proposed AOPP method. Numerical simulation of the asymptotic results shows that it performs better than NPPTF-QKD\cite{cui2019twin} at long distance regime and large noise regime.

\section{acknowledgement}
We thank Prof. Koashi for kindly providing us the source code of Ref.\cite{maeda2019repeaterless}. Hai Xu, Zong-Wen Yu and Cong Jiang contributed equally to this work. We acknowledge the financial support in part by The National Key Research and Development Program of China grant No. 2017YFA0303901; National Natural Science Foundation of China grant No. 11474182, 11774198 and U1738142.

\appendix

\section{Detail of $\tilde Z_0$-windows}\label{appendix:z0}
In the protocol, {\em they} use phase-randomized coherent states. This means, in the case that one and only one party (say, Alice) decides {\em sending}, she may actually sends a vacuum state. We define such time windows as $\tilde Z_0$-windows. The sent out state is identical to the case that Alice decides {\em not-sending}. This looks confusion on bit-value encoding. However, there is no self-inconsistency here: If one {\em only} looks at the sent-out state, not-sending is exactly the same with sending a vacuum. Since {\em they} use phase-randomized coherent states, a classical mixture of Fock states,  when she (he) decides {\em sending}, she (he) can then actually sends out a vacuum. But the bit values are determined by the $\emph{decisions}$ on sending or {\em not-sending} rather than the state sent out. Consider the cases {\em they} only send out vacuum. We can use a local state $|{\mathbb{S}}\rangle$ for the decision of {\em sending} and $|{\mathbb{N}}\rangle$ (orthogonal to $|{\mathbb{S}}\rangle$) for the decision of {\em not-sending}.  To Alice, if she decides {\em sending}, she can then actually send out vacuum $|{0}\rangle$; if she decides {\em not-sending} she also sends out vacuum. If we write out the state in the whole space, then $|{\mathbb{S}}\rangle |{0}\rangle$ represents bit value 1 and $|{\mathbb{N}}\rangle |{0}\rangle$ represents bit value 0. These two states are orthogonal.

Key rate formula. In Ref.\cite{wang2018twin}, we simply put all time windows with vacuum in transmission to tagged bits. But this is not necessary. The vacuum state in transmission for a time window when only one party decides {\em sending} is also un-tagged bits. Because in such a case Eve completely has no idea on which party makes the decision of {\em sending}. We actually can use
\begin{equation}\label{key01}
R=(n_1+n_0) - n_1 H(\bar e_1^{ph}) -f n_t H(E),
\end{equation}
to calculate the final key length, where $n_0$ is the number of {\em effective} $\tilde Z_0$-windows.
This can improve the performance of SNS a little bit than using the original key length formula Eq.(\ref{key0}).
We can also improve  Eq.(\ref{equ:keyrate1}) by
\begin{equation}
	N_f = n_{1}+n_0-n_1H(\bar e_1^{ph})-f[N_{0}H(E_{0})+N_{1}H(E_{1})],
\end{equation}

\section{Phase-flip error iteration formula (\ref{phase_odd}) in odd-parity error rejection}\label{appendix:odd}
Here we shall derive the phase-flip iteration formula for those un-tagged bits only.  Also, in the SNS protocol, there is no bit-flip error in the un-tagged bits.
For ease of presentation, we start from the virtual situation that initially Alice and Bob share $n_1$ entangled pair states which have no bit-flip error.
Obviously, SNS protocol is permutation invariant, and hence the de Finetti theorem \cite{renner2005security} applies. Asymptotically,
\begin{equation}
\rho_{AB}=\int_{\sigma} P(\sigma) \sigma^{\otimes n_1} d\sigma,
\end{equation}
where $n_1$ is the number of un-tagged bits.  Therefore, We can regard each pair state shared by Alice and Bob as an i.i.d. state. Consider the condition of zero bit-flip error for un-tagged bits in SNS protocol, here any pair-state is two dimensional.  For simplicity, we assume that any local state $|0\rangle$ ($|1\rangle$) corresponds to bit value 0 (1) for both Alice and Bob.
\begin{equation}\label{iid}
\begin{split}
\sigma=\cos^2\theta|{00}\rangle\langle{00}|+\sin^2\theta|{11}\rangle\langle{11}|\\
+\alpha \exp({-\mathbf{i}\beta})|{11}\rangle\langle{00}|+\alpha \exp({\mathbf{i}\beta})|{00}\rangle\langle{11}|.
\end{split}
\end{equation}
Considering the worst case for key rates, the phase-flip error rate is
\begin{equation}\label{eq:phase}
\begin{split}
e^{ph}(\sigma)=&\bar e^{ph}_1\\
=&tr(\hat M_{+}\sigma \hat M_{+}^{\dagger}+\hat M_{-}\sigma \hat M_{-}^{\dagger})\\
=&\frac{1}{2}(\cos^2\theta+\sin^2\theta-2\alpha\cos\beta)\\
=&\frac{1}{2}(1-2\alpha\cos\beta),\\
\end{split}
\end{equation}
where
\begin{equation}
\begin{split}
\hat M_{+}=|{+}\rangle_A\langle{+}|\otimes|{-}\rangle_B\langle{-}|\\
\hat M_{-}=|{-}\rangle_A\langle{-}|\otimes|{+}\rangle_B\langle{+}|,
\end{split}
\end{equation}
and
\begin{equation}
\begin{split}
|{+}\rangle=\frac{1}{\sqrt{2}}(|{0}\rangle+|{1}\rangle)\\
|{-}\rangle=\frac{1}{\sqrt{2}}(|{0}\rangle-|{1}\rangle).
\end{split}
\end{equation}
And $\cos\beta\le 1$ is always satisfied,
thus the phase-flip error rate for $\rho_{AB}$ can be expressed as
\begin{equation}\label{Eq:phase0}
\bar e^{ph}_1\ge  \frac{1}{2}(1-2\alpha),
\end{equation}
where the definition of $\bar e^{ph}_1$ is the same as before.

The parity check operations are taken on each sides of Alice and Bob, for notation clarity, we shall use $\otimes$ to divide the subspaces of Alice and Bob. For example, for a single-pair state in Eq.(\ref{iid}), we write it in the form:
\begin{equation}\label{iid1}
\begin{split}
\sigma=\cos^2\theta|{0}\rangle\langle{0}|\otimes |{0}\rangle\langle{0}|+\sin^2\theta|{1}\rangle\langle{1}|\otimes |{1}\rangle\langle{1}|\\
+\alpha \exp({-\mathbf{i}\beta})|{1}\rangle\langle{0}|\otimes |{1}\rangle\langle{0}|+\alpha \exp({\mathbf{i}\beta})|{0}\rangle\langle{1}|\otimes |{0}\rangle\langle{1}|.
\end{split}
\end{equation}
The density matrix of the two-pair state is $\sigma\cdot\sigma$. The un-normalized state of the survived pair after odd-parity error rejection can be represented by the
following formula\cite{kraus2007security}:
\begin{equation}
\tilde \rho_{odd} = (\hat M_A\otimes \hat M_B) (\sigma\cdot\sigma) ( \hat M^{\dagger}_{B}\otimes \hat M^{\dagger}_{A})
\end{equation}
where the conditional projection operators
\begin{equation}
\begin{split}
\hat M_{A}=|{0}\rangle\langle{01}|+|{1}\rangle\langle{10}|\\
\hat M_{B}=|{0}\rangle\langle{01}|+|{1}\rangle\langle{10}|.
\end{split}
\end{equation}
After normalization, we have the following bipartite single-pair state
\begin{equation}
\begin{split}
\rho_{odd}=&\frac{1}{2\cos^2\theta\sin^2\theta}[\cos^2\theta\sin^2\theta|{0}\rangle\langle{0}|\otimes |{0}\rangle\langle{0}|\\
+&\alpha^2|{0}\rangle\langle{1}|\otimes
|{0}\rangle\langle{1}|+\alpha^2|{1}\rangle\langle{0}|\otimes |{1}\rangle\langle{0}|\\ +&\cos^2\theta\sin^2\theta|{1}\rangle\langle{1}|\otimes |{1}\rangle\langle{1}|].
\end{split}
\end{equation}
This is the state for the survived pair if {\em they} find the odd parity values at each side.
According to Eq.(\ref{eq:phase}), the phase-flip error rate of the survived un-tagged bits after odd-parity check is
\begin{equation}
\begin{split}
e^{ph}_{odd}=\frac{1}{2}(1-\frac{\alpha^2}{\cos^2\theta\sin^2\theta}).
\end{split}
\end{equation}

Since $\cos^2\theta\sin^2\theta\le \frac{1}{4}$, we have
\begin{equation}
e^{ph}_{odd}\le \frac{1}{2}(1-4\alpha^2).
\end{equation}
Comparing with Eq.(\ref{Eq:phase0}), we have
\begin{equation}
\begin{split}
e^{ph}_{odd}\le& 2\bar e^{ph}_1(1-\bar e^{ph}_1).\\
\end{split}
\end{equation}
Thus the formula of the phase-flip error rate Eqs.(\ref{phase_odd}) is proved.

\section{Security Proof of AOPP}\label{appendix:AOPP}
Here we show the security of the proposed AOPP method in Sec.~\ref{sec.AOPP}.
Note that, the odd-parity sifting protocol with key-length Eq.(\ref{key:odd}) can be related with a virtual protocol with entangled pairs. Therefore its security is straightforward. However, in our last protocol, we use AOPP. This does not correspond to a virtual protocol of  entangled purification since we cannot guarantee to always obtain odd-parity results in the bipartite parity measurements. Therefore we need examine its security here. Our main idea is this: We can divide all odd-parity bit pairs from AOPP into two classes, the number of bit pairs in each class is not larger than the number of odd-parity pairs from odd-parity sifting which corresponds to a virtual entanglement purification protocol. Therefore, each class of bits in AOPP can be related to virtual entanglement purification protocol and hence final key from each class alone is secure. Since the mutual information of these two classes of bits is (almost) 0, then they are secure even the final keys of both classes are used.
Here we consider a virtual protocol first.\\

Prior to the pairing protocol, Bob has a sifted key $S$ containing $N_s$ bits.
He could obtain the information of $N_0$ and $N_1$, where the definitions of them are the same as Eq.(\ref{equ:keyrate1}).
He will create two sets of odd-parity bit pairs, set $\mathcal{W}_1$ contains $N_{W_1}$ pairs and set $\mathcal{W}_2$ contains $N_{W_2}$ pairs.
Also, we define string $\bar{S}_i$: from string $S$, after those bits which have been chosen to form pairs in set $\mathcal{W}_i$ are deleted, we obtain string $\bar{S}_i$.
Here $i$ can be 1 or 2. Clearly, string $\bar{S}_1$ contains $N_s-2N_{W_1}$ bits and string $\bar{S}_2$ contains $N_s-2N_{W_2}$ bits.

\noindent 1.
Bob estimates the number of odd-parity pairs he would obtain if he did random pairing, according to Eq.(\ref{eq:randompairing})
\begin{equation}
N_R=\frac{N_1N_0}{N_1+N_0}.
\end{equation}
According to Eq.(\ref{eq:activepairing}) in our main body text, if Bob did AOPP in $S$,
the number of odd-parity pairs he would obtain is
\begin{equation}
N_A=\min(N_1,N_0).
\end{equation}
We can see that
\begin{equation}
2N_R\ge N_A\ge N_R.
\end{equation}
On this basis, Bob decides two numbers $N_{W_1}$ and $N_{W_2}=N_A-N_{W_1}$, which satisfied
\begin{equation}
\begin{split}
&N_{W_1}\le N_R,\\
&N_{W_2}\le N_R.
\end{split}
\end{equation}

\noindent 2. Bob decides randomly on either taking operations stated by 2.1 or operations stated by 2.2 in the following:\\
\indent 2.1. 1).
Through randomly pairing of bits in $S$ for some times, Bob obtains $\lfloor N_s/2\rfloor$ pairs. He chooses $N_{W_1}$ odd-parity pairs and denotes them by set $\mathcal{W}_1$. He ignores the other pairs.
2).
Through AOPP to bits in set $\bar{S}_1$ (note that string $\bar{S}_i$ was defined earlier in the second paragraph of this section),
he obtains $N_{W_2}$ odd-parity pairs and denotes them by $\mathcal{W}_2$
\\
\indent 2.2. 1). Through randomly pairing of bits in $S$ for some times, Bob obtains  $\lfloor N_s/2\rfloor$ pairs. He chooses $N_{W_2}$ odd-parity pairs and denotes them by set $\mathcal{W}_2$, and ignores the other pairs.
2). Through AOPP to bits in set $\bar{S}_2$, he obtains $N_{W_1}$ odd-parity pairs. He denotes them by $\mathcal{W}_1$

\noindent 3. After the previous steps, Bob will obtain two odd-parity pairs sets $\mathcal{W}_1$ and $\mathcal{W}_2$. With such a setting, to anybody outside Bob's lab, each of these two sets could be obtained from random pairing by Bob. For convenience of description, we label the set obtained by random pairing with $\mathcal{R}$, and the set obtained by AOPP with $\mathcal{A}$.

\newtheorem{law}{Fact}
\begin{law}
	As long as Bob doesn't disclose his process, no one else will ever know where the set of pairs came from, and even if all the bits in the sifted key are announced, no one can tell which set was obtained by randomly pairing. This fact is the basis of the security proof.
\end{law}

\newtheorem{jury}{Lemma}
\begin{jury}
	If $\mathcal{W}_i$ was obtained by randomly pairing, that is $\mathcal{W}_i=\mathcal{R}$, the final key $r_i$ distilled from this set must be secure. Since $\mathcal{W}_i$ was obtained through random  pairing, the security of $r_i$ can be related to the virtual entanglement distillation. Therefore, the mutual information between $r_i$ and $\bar{S}_i$ must be negligible. Here $i$ can be 1 or 2.
\end{jury}

\noindent On the other hand, if set $\mathcal{W}_i$ is generated by AOPP, purely mathematically, we can also distill the final key $r_i$, and $r_i$ itself should be secure, and the mutual information between $r_i$ and $\bar{S}_i$ must be negligible. Otherwise Fact 1 will be violated. Then we have
\begin{jury}
	If $\mathcal{W}_i$ was obtained by AOPP, that is $\mathcal{W}_i=\mathcal{A}$, the final key $r_i$ distilled from this set must be secure  unconditionally, and the mutual information between $r_i$ and $\bar{S}_i$ must be negligible, for Eve has no way to tell which set is $\mathcal{R}$ and which one is $\mathcal{A}$. Here $i$ can be 1 or 2.
\end{jury}
We will prove Lemma 2 in detail later. So far we have three conclusions:
\begin{enumerate}
	\item key string $r_1$ itself is secure;
	\item key string $r_2$ itself is secure;
	\item the mutual information between $r_1$ and $r_2$ is negligible. These conclude that the final key $r_1\cup r_2$ is secure.
\end{enumerate}
Here the third conclusion can be obtained from Lemma 1 and Lemma 2, since either $r_1$ was distilled by $\mathcal{R}$ or $\mathcal{A}$, the mutual information between $r_1$ and $\bar{S}_1$ must be negligible, and $r_2$ was originally from $\bar{S}_1$.\\
Now we start to prove the Lemma 2. For clarity, we modify the expression of Fact 1. After obtained sifted key $S$, pair sets $\mathcal{W}_1$ and $\mathcal{W}_2$ were secretly generated by David. Then he hands both sets to Bob and disappears without providing any information. Bob has got these two sets and all the sifted keys, but no one except David knows which set was randomly paired. Then we have
\begin{law}
	After David disappears, no matter what Bob, Alice or Eve does, there's no way to know which set was randomly paired.
\end{law}
We denote the final key distilled from $\mathcal{A}$ by $r_A$, and the final key distilled from $\mathcal{R}$ by $r_R$. According to Lemma 1 we know that, Eve can't attack $r_R$ effectively. Now suppose that Lemma 2 is not tenable, that is to say, Eve has an effective means to attack $r_A$, which means,  Eve either gets more information  directly on $r_A$, or gets more information  after Bob announces $\bar{S}_A$. So Bob and Eve can work together to figure out which set is $\mathcal{A}$ and which set is $\mathcal{R}$:
\begin{enumerate}
	\item Bob announces $\bar{S}_1$;
	\item Eve uses the information $\bar{S}_1$ to attack $r_1$ and obtains the attack result t;
	\item Bob announces $r_1$;
	\item Based on $\bar{S}_1$ and $r_1$, Eve evaluated the attack effect, that is, the size of the information obtained by her attack on $r_1$, so as to judge whether $\mathcal{W}_1$ is $\mathcal{R}$ or $\mathcal{A}$.
\end{enumerate}
Since this conclusion is contrary to Fact 2, Lemma 2 has been proven.
It can be seen from the above discussion that the secret key extracted from $\mathcal{A}$ is secure, then the secret key extracted from $\mathcal{A}_1\cup\mathcal{A}_2$ is also secure. So Bob can simply form both of his sets $\mathcal{W}_1$, $\mathcal{W}_2$ by AOPP only, which is the protocol we proposed in Sec.\ref{sec.AOPP}. The security of the protocol is thus demonstrated.

\bibliography{refs}

\end{document}